\documentclass{article}
\usepackage{amssymb}

\usepackage[english]{babel}

\usepackage[letterpaper,top=2cm,bottom=2cm,marginparwidth=1.75cm]{geometry}

\usepackage[utf8]{inputenc}
\usepackage{csquotes}
\usepackage{amsmath}
\usepackage{graphicx}
\usepackage[colorlinks=true, allcolors=blue]{hyperref}
\usepackage{booktabs}
\usepackage{longtable}
\usepackage{float}
\usepackage{multirow}
\usepackage{authblk}
\usepackage{hyperref}

\usepackage{caption}

\usepackage{titlesec}

\usepackage[dvipsnames]{xcolor} 
\usepackage[normalem]{ulem}
\titlelabel{\thetitle.\quad}

\providecommand{\keywords}[1]
{
  \small	
  \textbf{\textit{Keywords---}} #1
}

\usepackage{soul}
\soulregister\cite7
\soulregister\ref7

\title{Critical role of phase-dependent properties in modeling photothermal sintering of LiCoO$_{2}$ cathodes}

\author[1,2,3]{Yang Hu}
\author[4]{Benoit Sklénard}
\author[3]{Wouter Vels}
\author[3]{Yaroslav E. Romanyuk}
\author[1,2,*]{Vladyslav Turlo}

\affil[1]{Laboratory for Advanced Materials Processing, Empa - Swiss Federal Laboratories for Materials Science and Technology, Feuerwerkerstrasse 39, 3602 Thun, Switzerland}
\affil[2]{National Centre for Computational Design and Discovery of Novel Materials MARVEL, Empa, Thun, Switzerland}
\affil[3]{Laboratory for Thin Films and Photovoltaics, Empa - Swiss Federal Laboratories for Materials Science and Technology, Ueberlandstrasse 129, D\"ubendorf, 8600, Switzerland}
\affil[4]{Univ. Grenoble Alpes, CEA, Leti, Grenoble F-38000, France} 

\affil[*]{Corresponding author: Vladyslav Turlo, vladyslav.turlo@empa.ch}

\begin{document}
\maketitle

\abstract{Photothermal (photonic) sintering crystallizes as-deposited amorphous LiCoO$_2$ (LCO) cathodes for solid-state thin-film batteries using millisecond, surface-localized heating. However, process design often relies on 1D models with phase-averaged, temperature-independent properties, which can mispredict peak temperatures and thermal damage margins. Here we develop a multiscale, data-driven framework that provides phase- and grain-size–resolved thermophysical inputs for stoichiometric LCO. We train an Allegro neural network potential with near-\textit{ab initio} accuracy, enabling Green–Kubo calculations of thermal conductivity for crystalline and amorphous phases. The low, weakly density-dependent conductivity of amorphous LCO motivates its use as an effective intergranular phase in a thin-interface model that reproduces observed grain-size–dependent thermal transport. Combined with measured wavelength-resolved optical properties in 1D multiphysics simulations, we show amorphous LCO absorbs more strongly and reaches higher peak temperatures than crystalline LCO; thus crystalline, constant-property models systematically overestimate safe operating windows.}


\maketitle
\newpage
LiCoO$_2$ (LCO) is a benchmark cathode material for solid-state batteries \cite{shao2003atomic,julien2019sputtered,ohzuku1994solid,Antolini2004159,huang1998electrochemical,zhang2019trace}, unlike conventional slurry-cast electrodes—typically made from already-crystalline LCO powder—thin-film LCO is often amorphous and requires a post-deposition crystallization step. Conventional furnace annealing (around 700 $^{\circ}$C for hours \cite{filippin2018ni}) increases manufacturing cost and energy consumption and imposes constraints on the choice of substrate and current collector \cite{filippin2018ni,filippin2017chromium}, limiting flexible and low-thermal-budget device architectures \cite{yim2020directly,clement2022recent}.
Photothermal (photonic) sintering, including flash-lamp annealing (FLA), intense pulsed light sintering, and UV “blacklight” flash-lamp sintering, offers a non-equilibrium alternative: millisecond-timescale light pulses can crystallize the cathode rapidly while limiting heating of underlying layers \cite{rebohle2019flash,chen2021photonic}. Because direct millisecond thermometry under intense illumination is challenging \cite{reichel2011temperature}, pulse design is commonly guided by simulated temperature profiles.

Many process models assume phase-averaged, temperature-independent optical and thermal properties, even though thin-film processing begins from the amorphous state and both absorption and heat removal can change substantially during crystallization and microstructural evolution. For example, the commercial software SimPulse is widely used for rapid process screening: it takes constant optical attenuation coefficient, thermal conductivity, and heat capacity for each layer in a stack and computes 1D transient temperature profiles for a prescribed pulse shape and energy \cite{guillot2012simulating}. However, such models typically neglect the wavelength dependence of reflectance/absorption and omit explicit temperature and grain-size dependence of thermal conductivity. Because photothermal sintering often operates close to crystallization, decomposition, or substrate-limiting thresholds, these simplifications can systematically bias predicted peak temperatures and safety margins, especially during early pulses when the film remains largely amorphous. 

Thermal conductivity is a particularly uncertain input for LCO: reported room-temperature values span nearly an order of magnitude (from $\sim$2.1 to $\sim$22 W/(mK) \cite{cho2020anisotropic,cho2014electrochemically,takahata2002thermal,mizuno2017thermoelectric}), elevated-temperature data are scarce \cite{mizuno2017thermoelectric}, and amorphous LCO lacks measurements. On the theoretical side, first-principles lattice-dynamics methods, such as quasi-harmonic phonon calculations and Boltzmann transport equation formalisms built from harmonic and higher-order force constants, are highly sensitive to the treatment of anharmonicity \cite{alfe2009phon,togo2023implementation,feng2020quantum} and are not applicable to amorphous phases, where the absence of long-range order precludes a conventional phonon description of heat transport. Equilibrium molecular dynamics (MD) based on the Green-Kubo formula is the golden standard that can treat both crystalline and amorphous heat transport, but direct \textit{ab initio} MD is prohibitively expensive at the required scales. Classical and machine-learning interatomic potentials (MLIPs) can overcome these limitations, yet existing LCO models often have limited accessibility and training coverage \cite{hart1998lattice,lee2018interatomic,sun2024diffusion}, making them unreliable at high temperature and for disordered or metastable structures (see Section S1). 
 
\begin{figure}[htbp!]
  \centering  \includegraphics[width=0.86\textwidth,clip,trim=0cm 0cm 0cm 0cm]{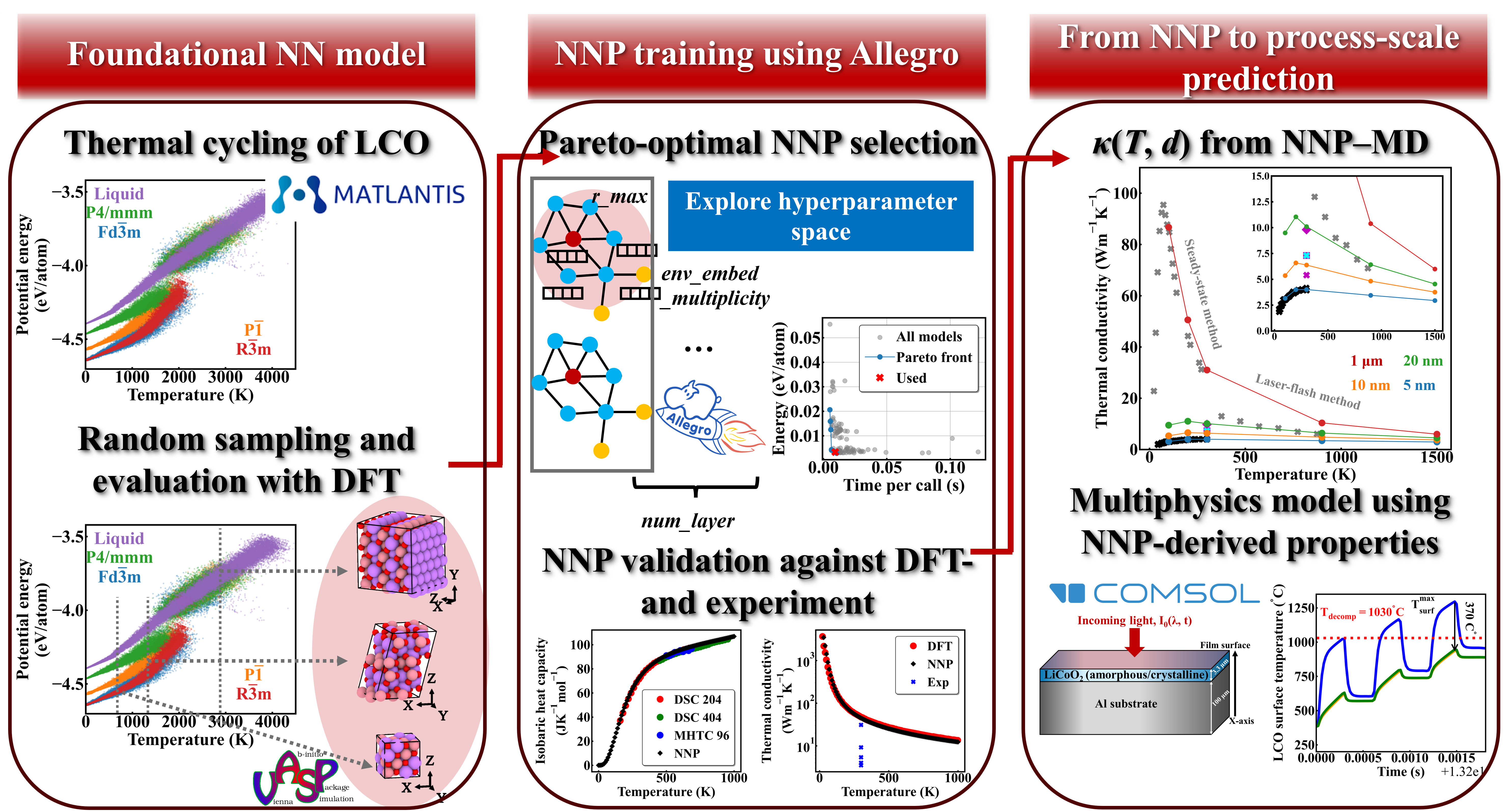}
  \caption{Detailed workflow of the study. (i) A foundational neural network potential, Matlantis \cite{Matlantis,takamoto2022towards}, is used for broad structure sampling of stoichiometric LiCoO$_2$ (LCO) across crystalline and disordered states. A benchmark set of 999 representative configurations is then recomputed with DFT+U+vdW (PBE+U with van der Waals corrections) \cite{perdew1996generalized,tolba2018dft,Grimme2010} to provide consistent reference energies, forces, and stresses. (ii) Using this dataset, an \textit{ab initio}-accurate neural network potential (NNP) is trained within the Allegro framework \cite{musaelian2023learning} and validated against DFT and experimental benchmarks for crystalline LCO (structure, elasticity, and vibrational properties). (iii) The validated NNP enables large-scale MD simulations to compute lattice thermal conductivity ($\kappa$) for both crystalline and amorphous LCO via the Green–Kubo approach. Grain-size effects in polycrystalline LCO are captured with a thin-interface grain-boundary model that treats amorphous LCO as an effective intergranular phase, yielding a grain-size–dependent effective $\kappa$. (iv) Finally, these phase- and grain-size–resolved thermophysical properties are combined with experimentally measured, wavelength-dependent optical inputs (reflectance and attenuation coefficient) in a 1D transient thermal transport model of LCO/Al stacks to predict pulse-driven temperature profiles. }
  \label{workflow}
\end{figure}

Here we test the hypothesis that phase-state-dependent optical absorption and thermal transport in LCO are sufficiently strong to alter temperature predictions and thus design decisions, and provide a route to supply these missing inputs (Fig.~\ref{workflow}). We curate a DFT+U+vdW benchmark set spanning crystalline and disordered configurations and train a dedicated Allegro neural network potential with near-\textit{ab initio} accuracy for energies, forces, and stresses (training and validation details are provided in Sections S3-S4). The potential reproduces structural, elastic, and vibrational properties of layered LCO in close agreement with experiment and prior first-principles studies, enabling Green–Kubo calculations of thermal conductivity for both crystalline and amorphous LCO. We then capture grain-size effects with a thin-interface model that treats amorphous LCO as an effective intergranular phase. Finally, these atomistically derived thermal properties are coupled with experimentally measured, wavelength-resolved optical data in 1D transient thermal transport simulations of LCO/Al stacks. We show that amorphous LCO absorbs more strongly and reaches higher peak temperatures than polycrystalline LCO under identical illumination, implying that crystalline, constant-property models can systematically overestimate safety margins and motivating microstructure-aware design of photothermal crystallization processes.

\section{Results}\label{sec2}
\subsection{Amorphous LCO as a proxy for as-deposited films and grain boundary regions}
\begin{figure}[htbp!]
  \centering
  \includegraphics[width=0.95\textwidth,clip,trim=0cm 0.0cm 0cm 0cm]{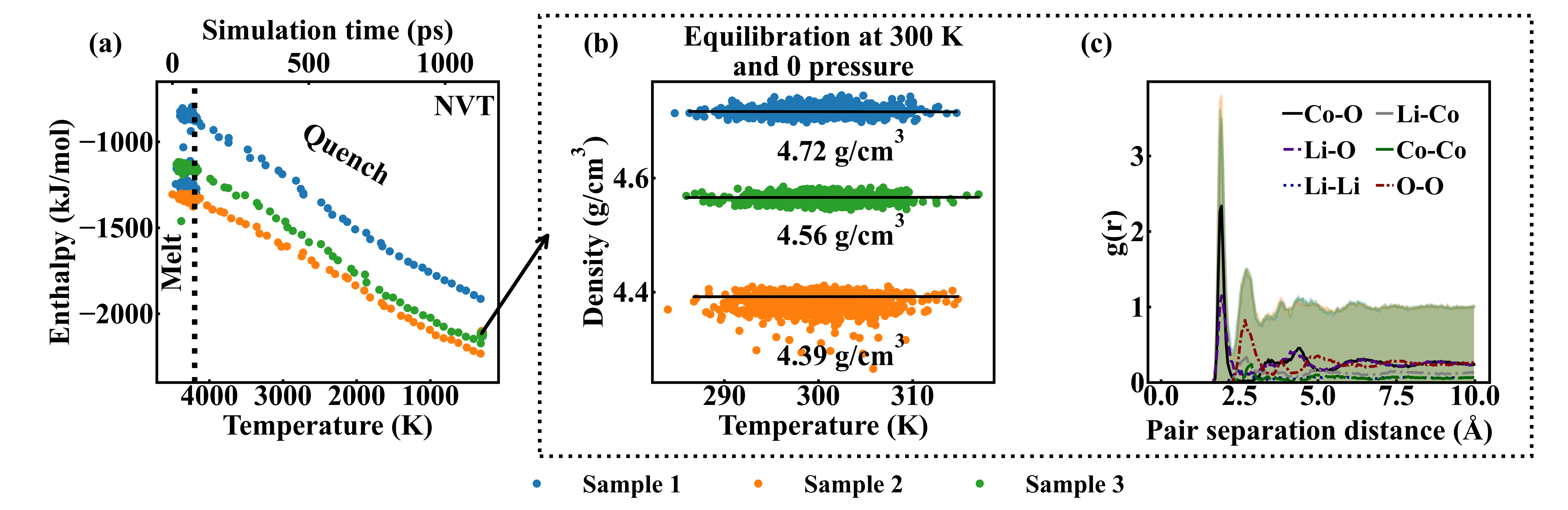}
  \caption{Amorphous LCO generated at different densities exhibits negligible structural differences. (a) Enthalpy and (b) density during thermal cycling under 0\%, 5\%, and 10\% applied uniform tensile strain. Each system is heated to 4300 K for 100 ps to induce complete amorphization, rapidly quenched to 300 K over 4 ps, and then equilibrated at 300 K in the isothermal-isobaric ($NpT$) ensemble for 1 ns to relax residual pressure. (c) Radial distribution functions $g(r)$ of the resulting amorphous structures. All three show a pronounced first peak at the characteristic metal–oxygen bond distance, dominated by Co–O correlations from local CoO$_6$ octahedra, with a weaker Li–O contribution at a similar distance reflecting the more variable Li coordination. A broader second peak at larger $r$ arises from mixed Li–Li, Li–Co, Co–Co, and O–O correlations, and its reduced amplitude/broadening indicates only weak medium-range order. At longer distances, $g(r)$ rapidly approaches 1, confirming the absence of long-range periodicity.}
  \label{cscs}
\end{figure}

We begin by quantifying heat transport in amorphous LCO—relevant to both as-deposited films and intergranular regions—and use it to define a grain-boundary conductivity, $\kappa_{gb}$. To this end, we generated a set of amorphous LCO configurations by the melt-quench approach under controlled cell strains, which produced amorphous structures spanning $\sim$87–93\% of the crystalline density (5.06 g/cm$^3$, see Table S1). Radial distribution functions confirm the expected amorphous character—well-defined short-range order with no long-range periodicity—and are nearly indistinguishable across this density range (Fig.~\ref{cscs}c). 

Using these configurations, we computed the lattice thermal conductivity at 300 K via equilibrium MD using the Green–Kubo formalism with the Allegro potential (Methods). Despite the density variation, the amorphous-phase thermal conductivity remains in a narrow range of 1.19-1.32 W/(mK) with no systematic trend with density. This indicates that thermal transport in amorphous LCO is dominated by strong local disorder rather than modest changes in mass density. These values correspond to an approximately 95.89\% reduction relative to the room-temperature thermal conductivity of monocrystalline LCO (see Fig.~S7(a)). In the subsequent grain-size model, we therefore treat the thermal conductivity of the amorphous phase as effectively density-independent and use a single representative value for the grain-boundary conductivity, $\kappa_{gb}$. For reference, Cho et al. \cite{cho2014electrochemically} reported a thermal conductivity of 2.1 W/(mK) for as-deposited LCO films, which were described as largely amorphous. 

\subsection{Resolving thermal conductivity across temperatures and grain size}
To incorporate microstructural effects in continuum-scale photothermal simulations, we model the effective thermal conductivity of polycrystalline LCO using the thin-interface grain-boundary framework of Badry and Ahmed \cite{badry2020new}. In this model, grain boundaries are treated as a finite-thickness intergranular phase of width $l$ with its own thermal conductivity $\kappa_{gb}$, rather than as a zero-thickness interface. The resulting analytical expression for the effective thermal conductivity is

\begin{equation}
    \kappa_{eff}=\frac{(d+l)}{\frac{d}{\kappa_{grain}}+\frac{l}{\kappa_{gb}}}
\end{equation}

where $d$ is the grain size, $\kappa_{grain}$ is the conductivity of the crystalline grain interior, and $\kappa_{gb}$ characterizes intergranular transport. We set $\kappa_{gb}$ to the amorphous LCO thermal conductivity obtained from Green–Kubo MD, 1.3 K/(mK) at 300 K; this value is nearly insensitive to the modest density variation of our amorphous samples. For $\kappa_{grain}(T, d)$, we start from the single-crystal conductivity $\kappa_{bulk}(T)$ computed by Green–Kubo MD and include finite-size suppression in small grains using a mean-free-path correction (details in Section~S5). The only remaining parameter is the grain-boundary width $l$, which we calibrate using the time-domain thermoreflectance (TDTR) measurements of Cho \textit{et al.} \cite{cho2014electrochemically} for annealed LCO films with reported grain size ($\sim$10 nm) and thermal conductivity at 300 K, yielding $l\approx$1.8 nm. We then hold $l$ fixed for all temperatures and grain sizes. 

\begin{figure}[htbp!]
  \centering
  \includegraphics[width=0.7\textwidth,clip,trim=0cm 0.0cm 0cm 0cm]{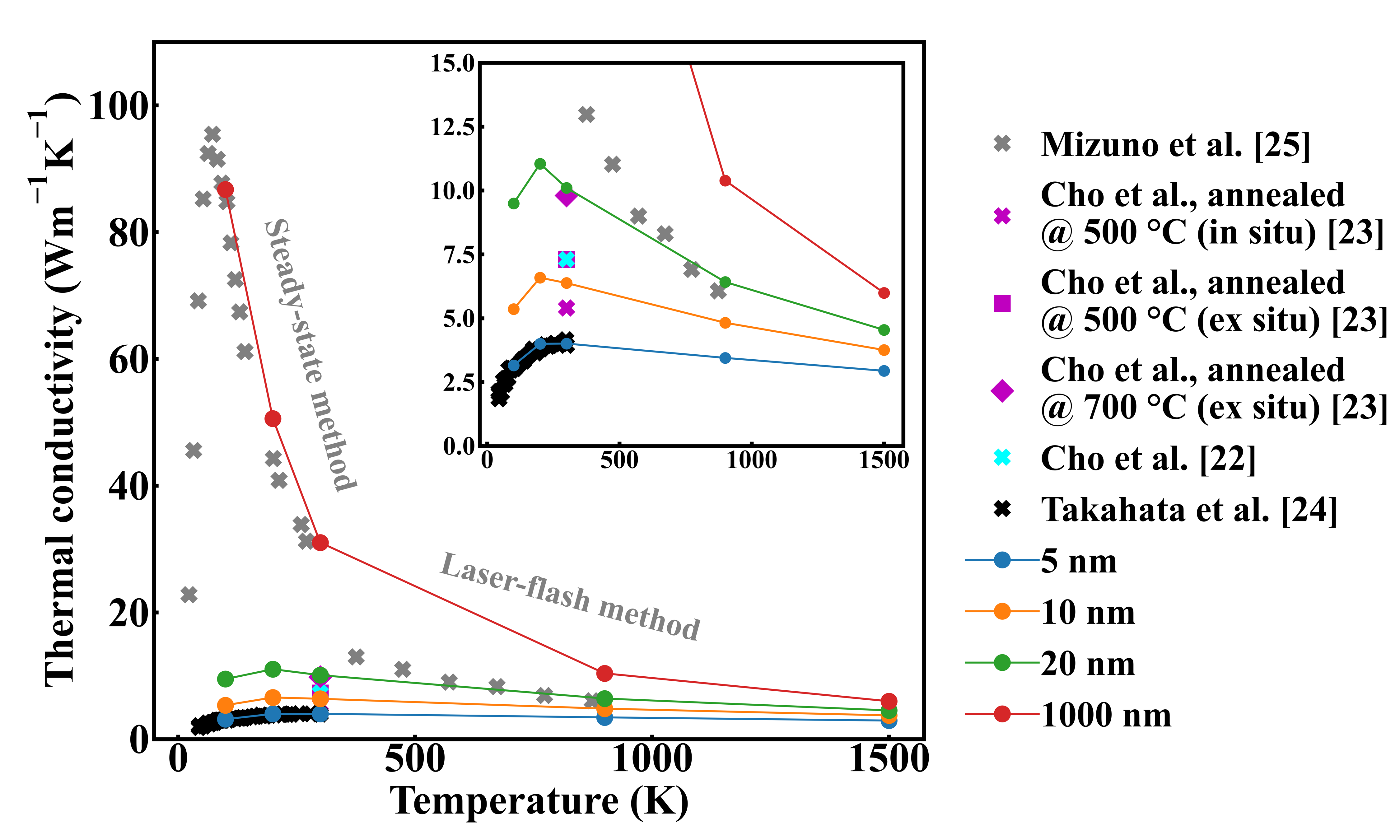}
  \caption{Grain-size dependence of the thermal conductivity at 100, 200, 300, 1000, and 1500 K. Experimental data are taken from Refs.~\cite{mizuno2017thermoelectric} (gray crosses), \cite{cho2020anisotropic} (cyan crosses), \cite{cho2014electrochemically} (magenta markers), and \cite{takahata2002thermal} (black crosses). For Ref.~\cite{cho2020anisotropic}, the plotted value is the orientation-averaged conductivity, 1/3(2$\times\kappa_{\parallel}+\kappa_{\perp}$).}
  \label{grain_tc}
\end{figure}

Fig. \ref{grain_tc} shows the resulting $\kappa_{eff}(T,d)$ together with representative experimental data. At low temperatures, grain-boundary scattering dominates, and $\kappa_{eff}$ increases strongly with grain size, whereas at higher temperatures Umklapp scattering reduces the overall conductivity and largely suppresses grain-size sensitivity, leading to convergence across $d$. With a single fitted $l$, the model captures the main trends across disparate datasets, including the strong reduction in $\kappa$ in nanocrystalline films and the weaker microstructural dependence at elevated temperature.

In Ref.~\cite{cho2014electrochemically} (magenta markers), Cho \textit{et al.} used TDTR to quantify how microstructure and lithiation affect the room-temperature thermal conductivity of LCO thin films. Their as-deposited film, prepared by room-temperature RF sputtering, was largely amorphous and exhibited a low 2.1 Wm$^{-1}$K$^{-1}$. Upon annealing at 500 $^{\circ}$C in air for 1 h, the film crystallized into a nanocrystalline microstructure with grain sizes $\sim$10 nm and $\kappa$ increased to 7.3 Wm$^{-1}$K$^{-1}$. Further annealing at 700 $^{\circ}$C produced a highly crystalline film with grain sizes of several tens of nanometers and $\kappa$ rose to 9.8 Wm$^{-1}$K$^{-1}$. For the 500 $^{\circ}$C-annealed film, Cho \textit{et al.} also performed \textit{in situ} TDTR during electrochemical cycling, reporting 5.4 Wm$^{-1}$K$^{-1}$ for fully lithiated LCO—lower than the \textit{ex situ} value of 7.3 Wm$^{-1}$K$^{-1}$. They attributed this difference to the measurement geometry and sampling region: the \textit{in situ} configuration probes the Al/LCO interface within the liquid cell, whereas \textit{ex situ} TDTR probes the top surface after deposition of a fresh Al transducer; interface quality, electrolyte exposure, and near-collector heterogeneity can therefore reduce the apparent $\kappa$. Because Cho et al. reported both grain size and corresponding thermal conductivity, these data were used to calibrate the only unknown parameter in our grain-size model—the grain-boundary thickness, $l$, as mentioned above. Since we use the average of the two reported $\kappa$ values for the 500 $^{\circ}$C-annealed film to determine the grain-boundary thickness, our predicted room-temperature $\kappa$ at $d$=10 nm naturally falls between the two corresponding data points. The 700 $^{\circ}$C-annealed film aligns closely with the model prediction at 300 K for $d\sim$20 nm (Fig.~\ref{grain_tc}).

In Ref.~\cite{mizuno2017thermoelectric} (gray crosses), polycrystalline LiCo$_{1-x}$M$_x$O$_2$ (M = Cu, Mg, Ni, Zn) was synthesized by solid-state reaction followed by spark plasma sintering. Thermal conductivity was measured from 300–873 K using the laser-flash method under vacuum, while values below 300 K were obtained by a steady-state heat-flow method \cite{fujishiro1994anisotropic}. Our model prediction with $d\sim$1 \textmu m agrees well with the low-temperature steady-state data (below 300 K), but deviates from the high-temperature laser-flash results (above 300 K). Notably, the dataset shows a sharp drop across 300 K when switching measurement techniques; such discontinuities can reflect method-dependent systematic biases and sample-dependent effects. In particular, at elevated temperatures, the accuracy of the laser-flash method—especially for low-$\kappa$ oxides—can be compromised by radiative heat losses, porosity, non-uniform laser illumination, and sample inhomogeneity \cite{corbin2002thermal,Hicham2019,lunev2020decreasing}. If these effects are not rigorously accounted for, they tend to bias the extracted diffusivity and thus $\kappa$ to lower values than those obtained from steady-state techniques, which determine heat flow under near-equilibrium conditions. 

In Ref.~\cite{takahata2002thermal} (black crosses), polycrystalline LCO synthesized by conventional solid-state reaction was measured by a steady-state method from 15–280 K. Our model reproduces these low-temperature data closely using an effective grain size $d\sim$5 nm. Takahata \textit{et al.} analyzed their measurements with a Debye–Callaway model \cite{callaway1959model,callaway1960effect} including Umklapp and boundary scattering (and neglecting point-defect scattering), and obtained a fitted boundary-limited length scale of $\sim$3 nm. Our grain-size model likewise neglects additional scattering channels (e.g., point defects, dislocations, and strain fields). When these mechanisms are weak or secondary, the fitted boundary term can still be interpreted in terms of a physically meaningful grain size. However, when they contribute appreciably, their effects are inadvertently folded into the fitted “boundary” term, and the inferred nanometer-scale grain sizes become an effective scattering length rather than a literal microstructural dimension. In this case, the fitted length aggregates scattering from grain boundaries and other extended defects, which limits quantitative interpretation of the extracted grain size even if the model reproduces the overall $\kappa(T)$ trend. 

Finally, in Ref.~\cite{cho2020anisotropic} (cyan crosses), epitaxial LCO films were prepared on sapphire by reactive solid-phase epitaxy followed by topotactic ion exchange, enabling TDTR measurement of anisotropic thermal conductivities. The reported orientation-averaged $\kappa\approx$7.3 Wm$^{-1}$K$^{-1}$, is consistent with the 500 $^{\circ}$C-annealed film in Ref. \cite{cho2014electrochemically}. Overall, these comparisons show that a simple analytical grain-size model—anchored by an atomistically derived choice of $\kappa_{gb}$, can rationalize a wide range of experimentally reported thermal conductivities for LCO within a unified framework. While the extracted “grain sizes” should be interpreted as effective scattering lengths rather than literal microstructural dimensions, the model still captures the correct trends with temperature and microstructure and reproduces low-temperature data semi-quantitatively. In this sense, it provides practical, microstructure-aware $\kappa(T,d)$ inputs for device-scale photothermal simulations, grounded in atomistic thermal-transport calculations.

\subsection{Design guidelines for photothermal sintering from multiphysics simulations}
To provide general guidance for selecting photothermal sintering parameters in experiments, one-dimensional transient temperature profiles T(x,t) were simulated using the COMSOL Multiphysics© software \cite{multiphysics1998introduction}. The model solves heat conduction in an LCO/Al stack under a time- and depth-dependent volumetric heat source derived from the lamp emission spectrum, pulse waveform, and the measured, wavelength-resolved optical response of LCO. During processing, LCO films undergo transformation from amorphous to polycrystalline, accompanied by grain growth and potential changes in density and stress state. Rather than attempting to capture this full coupled evolution (phase transformation kinetics, mass transport, grain growth, and thermomechanical stress), we isolate a central materials-physics question: how strongly do phase-dependent optical absorption and thermal transport (amorphous vs polycrystalline) alter transient heating predictions, and therefore process windows? This is particularly important because photothermal treatments often operate near crystallization, decomposition, or substrate-limited thresholds; small changes in absorptance or thermal conductivity can determine whether the film enters a damage regime within a millisecond pulse train. 

\begin{figure}[htbp!]
  \centering
  \includegraphics[width=0.95\textwidth,clip,trim=0cm 0.0cm 0cm 0cm]{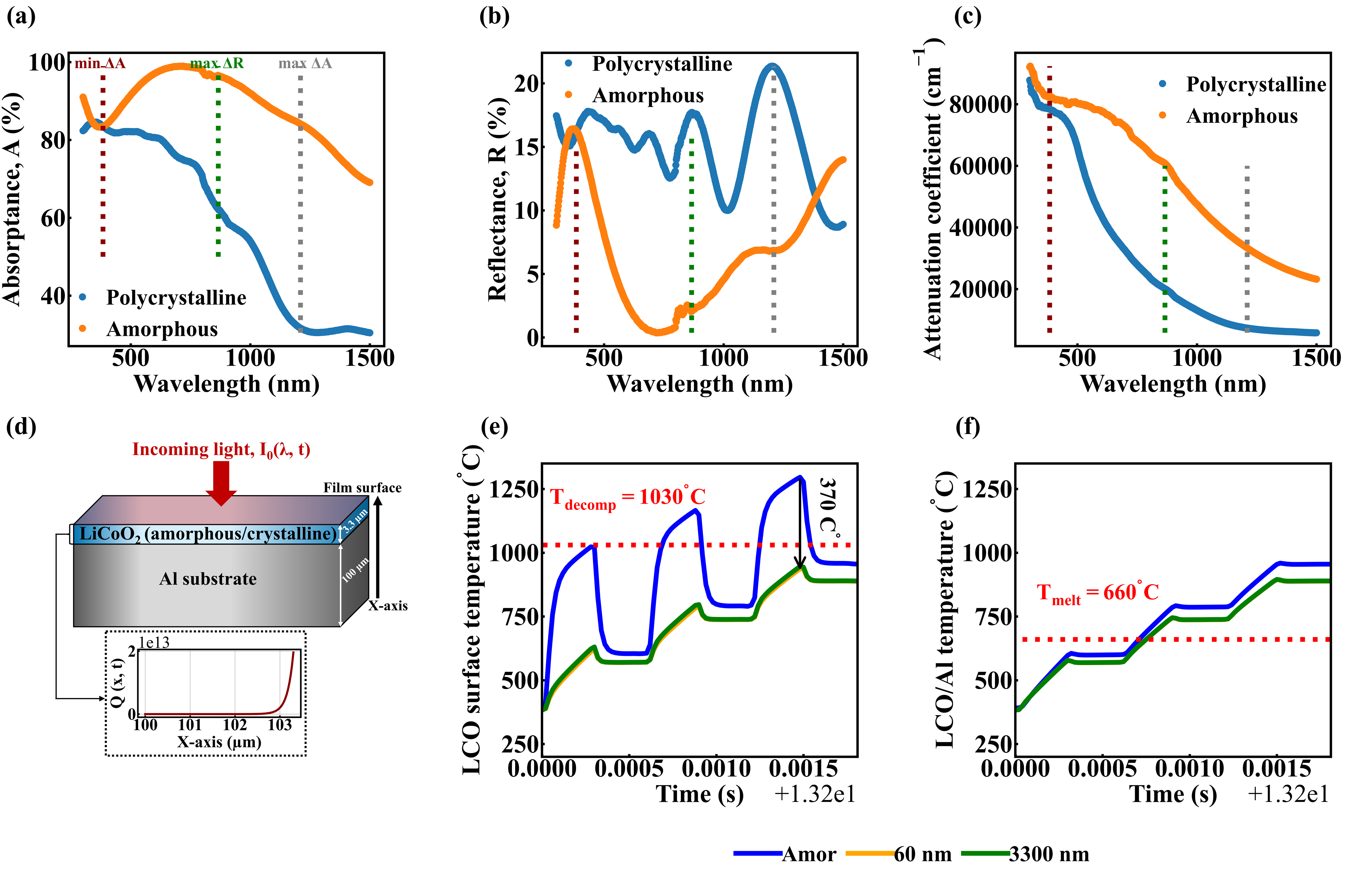}
  \caption{Multiphysics simulations, informed by measured optical and computed thermal properties, show that amorphous LCO films reach higher surface temperatures than crystalline films across the investigated spectrum. Measured wavelength-dependent (a) absorptance (A), (b) reflectance (R), and (c) attenuation coefficient ($\alpha$) of amorphous and polycrystalline LCO films. The vertical dashed lines mark the wavelengths corresponding to the minimum and maximum differences in absorptance ($\Delta$A) between the two phases, as well as the maximum difference in reflectance ($\Delta$R). (d) Schematic illustration of the COMSOL simulation setup: a two-layer stack consisting of a thin LCO film (amorphous or polycrystalline) on an Al substrate. The incident light enters from the top surface, and absorption in the LCO is modeled as a wavelength-resolved, depth- and time-dependent volumetric heat source Q(x,t) applied throughout the LCO layer (attenuating with depth according to $\alpha(\lambda)$); heat then conducts into the Al substrate. (e) Simulated LCO surface temperature for amorphous and polycrystalline films (grain sizes: 60 and 3300 nm) during a single experimental FLA pulse (850 V, 1800 \textmu s), using measured wavelength-dependent optical properties. The red dotted line marks the reported onset of LCO decomposition, $\sim$1300 K \cite{jankovsky2016thermodynamic}. (f) Simulated LCO/Al interface temperature for the same amorphous and polycrystalline films under identical pulse conditions. The red dotted line indicates the melting point of Al. The red dotted line indicates the melting point of Al.}
  \label{comsol}
\end{figure}

The wavelength-dependent optical properties of amorphous (as-deposited) and polycrystalline (annealed at 500 $^{\circ}$C for 2 hours) LCO were obtained experimentally from films deposited on borosilicate glass. Reflectance (R) and transmittance (T) were measured using a UV–Vis–NIR spectrophotometer, from which absorptance was computed as A=1-R-T, and the attenuation coefficient ($\alpha$) was derived using the Beer–Lambert relation. Across the measured spectral range (300–1500 nm), amorphous LCO exhibits markedly higher absorptance and attenuation than polycrystalline LCO (Fig.~\ref{comsol}a–d), exceeding 90\% near 500 nm, whereas the polycrystalline film shows lower absorptance that gradually decreases with wavelength. The minimum difference in absorptance ($\sim$0.03\%) occurs near 380 nm, while the maximum difference ($\sim$52.5\%) appears around 1200 nm. Similarly, the attenuation coefficient of the amorphous film remains consistently higher than that of the polycrystalline sample, with a maximum difference of 2$\times$10$^4$ cm$^{-1}$ observed near 860 nm. In addition, the polycrystalline LCO film exhibits a higher reflectance with pronounced oscillations across the spectrum, whereas the amorphous film shows a generally lower and smoother reflectance over the entire 300–1500 nm range. The difference in reflectance between the two films is modest at short wavelengths (near the minimum $\Delta$A), but increases toward the near-infrared and reaches a maximum around $\sim$870 nm. This reduced reflectance of the amorphous film, combined with its larger attenuation coefficient, is consistent with the significantly higher absorptance observed for the amorphous phase.

All optical properties were measured at room temperature, and in multiphysics simulations we treat R(T) and $\alpha$(T) as temperature independent and use the 300 K spectra as a first-order approximation. For many transition-metal oxides, the temperature-induced changes in refractive index and extinction coefficient over the range relevant for photonic sintering are typically modest compared to the strong contrast between amorphous and crystalline phases; thus, this approximation is reasonable. In this framework, the absorbed power density in the LCO film scales approximately with (1-R)$\alpha$, so the smaller reflectance and larger attenuation coefficient of amorphous LCO should lead to stronger and more surface-localized heating than in the polycrystalline film under identical illumination. Conversely, the polycrystalline film, with higher reflectance and lower $\alpha$, absorbs less energy and distributes it over a larger depth. Although the exact temperature profiles would be slightly modified by a full temperature-dependent optical model, the qualitative implication is robust: amorphous LCO couples more incident energy into the film and deposits it more surface-locally than polycrystalline LCO under identical illumination. This enhanced absorption is consistent with disorder-induced band-tail and defect-related states that broaden optical transitions in amorphous oxides \cite{ferlauto2002analytical,kaiser2021universal,andronic2020black}, see Section S6 for detailed discussion.

Accounting for structure-dependent optical properties is therefore critical for accurately predicting temperature distributions and thermal gradients during photonic sintering processes~\cite{rebohle2019flash}.
To the best of our knowledge, this work represents the first detailed optical characterization of amorphous and crystalline LCO thin films in the context of thermal processing applications. While the electronic and electrochemical properties of LCO have been extensively studied for battery applications, the wavelength-dependent optical properties—particularly the absorption coefficient—have not been previously reported in the literature. The substantial difference in optical absorption between amorphous and crystalline phases highlights the importance of considering structural state when modeling light-matter interactions in LCO thin films.

For the multiphysics simulations, the wavelength dependence of the optical response is included by estimating the volumetric heat source ($Q(x,t)$) as,
\begin{equation}
    Q(x,t)=\int_{\lambda} (1-R(\lambda))I_0(\lambda,t)(\int_x e^{-\alpha(\lambda) x}dx) d\lambda
\end{equation}
where $\lambda$ is the wavelength, $\alpha(\lambda)$ and $R(\lambda)$ are the wavelength-dependence of attenuation coefficient and reflectance, $x$ is the position in the film, and $I_0(\lambda,t)$ is the incident light intensity as a function of wavelength and time. The incident intensity $I_0(\lambda,t)$ is constructed from the measured lamp spectrum and pulse waveform (Section~S7).
A schematic of the COMSOL simulation setup is shown in Fig. \ref{comsol}(d). The model consists of a two-layer stack with a 3.3 \textmu m-thick LCO film (amorphous or crystalline) deposited on a 100 \textmu m-thick Al substrate, with the heat source applied to the LCO layer. The transmitted heat then propagates into the Al layer according to the Beer–Lambert law. Aluminum is a practical substrate/current collector due to its high electrical conductivity, low cost, roll-to-roll compatibility, and high in-plane thermal conductivity that spreads heat and reduces hot spots.
The thermal conductivity of amorphous LCO is assumed to be only weakly temperature dependent, whereas for polycrystalline LCO an inverse temperature dependence of $\kappa$ is implemented based on equilibrium MD using the Green–Kubo formalism (see Fig. S10). This behavior reflects the increasing dominance of Umklapp phonon scattering at elevated temperatures. For heat capacity, we use the Dulong–Petit limit (3NR) for both phases; this approximation is supported by the near-linear $\Delta$H–$\Delta$T relation from our atomistic calculations (Fig.~S12). Heat dissipation is allowed at the top LCO surface and at the bottom of the Al layer, following the same boundary conditions as in \cite{chen2021flash}.

Except for the amorphous phase, we simulated polycrystalline LCO with an effective grain size of 60 nm—consistent with RF-sputtered LCO films annealed at 500 $^\circ$C for 2 h reported in \cite{liao2004lithium}—and an upper-bound grain size of 3.3 \textmu m (equal to the film thickness), assuming a temperature-independent grain size throughout the flash sequence. This simplification neglects grain nucleation and growth during annealing, but it allows us to explore how a reasonable range of $\kappa$ values impact the transient temperature response.
Figure \ref{comsol}(e) shows the simulated film-surface temperature during a single experimental FLA pulse sequence at 850 V (total duration $\approx$ 1800 \textmu s, see Fig. S9(b)), using the pulse shape together with the full emission spectrum shown in Fig. S9(a). The temperature exhibits a series of rapid rises and partial cool-downs that follow the micro-pulses of the lamp. The surface temperature of the amorphous film climbs to several hundred $^\circ$C within a few hundred microseconds, reflecting both the extremely rapid heating characteristic of FLA and the combination of higher absorptance and much lower thermal conductivity of the amorphous phase. By contrast, polycrystalline LCO with effective grain sizes of 60 nm and 3.3~\textmu m remains within a much narrower temperature band: smaller grains (lower $\kappa$) reach slightly higher peaks, but the crystalline films still stay several hundred degrees cooler than the amorphous one, with a maximum temperature difference of about 370 $^{\circ}$C. Notably, for this tested configuration (850 V pulse, 3.3 \textmu m film thickness), the two polycrystalline grain-sizes produce relatively similar surface-temperatures compared with the much larger amorphous–crystalline separation, indicating that the phase-dependent optical response (via R($\lambda$) and $\alpha$($\lambda$)) is the dominant driver of surface heating under these conditions. 
Figure~\ref{comsol}(f) reports the corresponding LCO/Al interface temperature, which remains substantially lower than the surface temperature and is less sensitive to the optical contrast, reflecting the combined effects of heat diffusion through the film and the high thermal conductivity/heat capacity of the Al substrate. 

For the amorphous case, the surface temperature exceeds the reported onset of LCO decomposition \cite{jankovsky2016thermodynamic}, and the interface temperature of both amorphous and polycrystalline film exceeds the Al melting point. We therefore use the surface temperature—maximal across the LCO thickness—to assess film degradation risk, and the interface temperature—maximal within the substrate thickness—to provide a conservative indicator of potential substrate melting or softening.
Importantly, exceeding these thresholds in a millisecond pulse does not necessarily imply immediate catastrophic failure. Whether a pulse condition is practically usable depends on how rapidly crystallization proceeds and how quickly the resulting changes in optical/thermal properties drive the system into a self-limiting, lower-temperature regime. In this context, the relevant risk metric is the duration spent above critical temperatures rather than the peak temperature alone. As crystallization and grain coarsening progress, the optical response may also evolve; for example, increasing grain size reduces the volume fraction of disordered intergranular regions, which could decrease absorptance and further suppress heating. Overall, the large separation between amorphous and polycrystalline responses underscores the need to incorporate phase-dependent optical and thermal inputs when translating pulse parameters into quantitative experimental design rules.

\section{Discussion}
The results clarify when detailed atomistic information is actually necessary for modeling photothermal processing of LCO. In regimes where the film is already crystalline and far from decomposition, phase-averaged, weakly temperature-dependent properties may be sufficient for empirical tuning. However, our calculations show that this approximation breaks down in the early stages of processing, when the cathode is still amorphous: in this regime, thermal transport and light–matter interaction are both strongly phase dependent. The low, nearly density-independent thermal conductivity of amorphous LCO and its enhanced optical absorption jointly amplify temperature rise under pulsed illumination, tightening the margin between “useful” crystallization and the onset of decomposition. In other words, the initial amorphous state is not a small perturbation to a crystalline effective medium, but a qualitatively different thermal and optical regime that must be explicitly accounted for in models that aim to be predictive rather than purely empirical.

Within this context, the contrast between amorphous and crystalline LCO becomes the central design lever rather than a detail. The simulations show that, under identical pulse conditions, amorphous films consistently reach higher surface temperatures and cross decomposition-relevant thresholds earlier than polycrystalline films, while crystalline films of different grain sizes cluster within a relatively narrow thermal response band. This explains why models that implicitly assume a crystalline effective medium and constant $\kappa(A)$ can systematically overestimate safe operating windows when the process is actually limited by the amorphous phase. It also rationalizes the “self-limiting” character of flash-lamp annealing: once crystallization proceeds and $\kappa$ increases while absorptance drops, subsequent pulses couple less energy into the film and peak temperatures naturally saturate. From a process-design perspective, this suggests that the relevant question is not only “What peak temperature is reached?” but “How long does the film spend in the high-absorption, low-$\kappa$ amorphous state under a given pulse train?”—a quantity that our $\kappa(T,d)$ + optical-property–informed maps are explicitly designed to constrain.

Going forward, a key challenge is to move from static “phase-tagged” properties to fully crystallization-aware transport models. In the present work, amorphous and crystalline LCO are treated as distinct phases, with effective thermal conductivities and optical properties assigned to each. A natural next step is to couple the thermal model to a crystallization kinetics description—so that the local crystalline fraction evolves in time and space—and to update transport coefficients accordingly. In analogy to our thin-interface treatment of grain-boundary thermal conductivity, microstructure-resolved mixing rules could be used to construct effective $\kappa(T, f_{\mathrm{cryst}})$ and, equally importantly, effective optical properties $A(\lambda, f_{\mathrm{cryst}})$, $R(\lambda, f_{\mathrm{cryst}})$, and $\alpha(\lambda, f_{\mathrm{cryst}})$ that evolve as the film crystallizes. Such a framework would explicitly capture the feedback loop between crystallization kinetics, changing absorption, and self-limiting temperature rise under pulsed illumination.

Although demonstrated here for stoichiometric LCO on Al, the overall strategy—foundational-potential sampling, targeted DFT+U+vdW labeling, NequIP/Allegro-based potential training, Green–Kubo transport, and wavelength-resolved multiphysics coupling—is not specific to this chemistry or geometry. The same workflow can, in principle, be applied to garnet and perovskite electrolytes, composite cathodes, printed conductors, and functional oxides processed on temperature-sensitive, often polymeric, substrates in flexible electronics and other thin-film devices. In this broader context, machine-learned interatomic potentials provide the missing bridge between microstructure-dependent thermophysical properties and device-scale photothermal simulations, enabling materials-informed process windows for a wide range of photonic sintering and rapid-annealing technologies.

\section{Methods}
\subsection{Molecular dynamics simulations}
Molecular dynamics (MD) simulations were performed using two frameworks: the Large-scale Atomic/Molecular Massively Parallel Simulator (LAMMPS) package \cite{Plimpton1995}, and the Atomistic Simulation Environment (ASE) \cite{larsen2017atomic}, integrated with the NequIPCalculator module for deploying trained neural network potentials (NNPs). LAMMPS enables efficient parallelization through message-passing and spatial domain decomposition, allowing simulations to scale well across many compute nodes. ASE, written in Python, provides a user-friendly interface for setting up, managing, and analyzing atomistic simulations. It supports various tasks including structure optimization, MD, constrained simulations, and nudged elastic band (NEB) calculations.

To model atomic interactions, two interatomic potentials are employed. The first is Matlantis \cite{takamoto2022towards,Matlantis}, a foundational neural network potential trained on over 60 million density functional theory (DFT) calculations covering 96 elements. The second potential used in this work was trained using the Allegro model, a high-performance, E(3)-equivariant graph neural network optimized for hybrid CPU+GPU architectures.

Post-processing and analysis of MD trajectories are performed using OVITO \cite{Stukowski2010}, a powerful tool for visualizing and analyzing atomistic data. Local atomic environments are characterized using the polyhedral template matching (PTM) method \cite{larsen2016robust}, which assigns a crystalline structure and orientation quaternion to each atom. Crystallinity—whether amorphous or nanocrystalline—is further analyzed using radial distribution functions (RDFs) within OVITO. The number of grains and grain sizes are estimated using the automatic graph clustering algorithm embedded in OVITO, with a minimum grain size equal to 20 atoms. 

\subsubsection{Melting and cooling simulations on LCO structures using the universal neural network potential, Matlantis} \label{LCO_structures}
Four distinct crystal structures of LiCoO$_{2}$ are selected for MD simulations using Matlantis. These structures, illustrated in Fig. S1, are obtained from the Materials Project (MP) database \cite{jain2020materials}. The energetically most stable structure, MP\_22526, corresponds to the layered rhombohedral phase (space group $R\overline{3}m$). For this structure, an orthorhombic simulation cell is constructed by aligning the X-axis along the [100]-direction, the Y-axis along the [120]-direction, and the Z-axis along the [001]-direction. This simulation box is a 5$\times$6$\times$1 supercell containing 360 atoms. For the remaining structures, the simulation boxes are constructed as follows: MP\_849273 uses the original 64-atom unit cell from the MP database; MP\_867664 is modeled with a 2$\times$2$\times$2 supercell (160 atoms); and MP\_1222334 with a 3$\times$3$\times$3 supercell (108 atoms).

To simulate melting, each of the four crystalline structures is gradually heated from 0.01 K to 3000 K over 300 ps under the isothermal-isobaric ($NpT$) ensemble at zero pressure. Temperature and pressure control are applied using the Nose–Hoover thermostat and barostat \cite{shinoda2004rapid}, with a time step of 0.1 fs to ensure both numerical stability and physical fidelity. This setup is consistently applied in all subsequent MD simulations using Matlantis.

Further simulations are carried out on the liquid phase obtained by melting the 360-atom layered LiCoO$_2$ structure (MP\_22526). The system is cooled from 4000 K to 0.01 K over 400 ps. Due to the rapid quenching, crystallization is kinetically hindered, resulting in the formation of an amorphous (glassy) phase. This amorphous configuration at 0.01 K is then subjected to additional simulations: one under +1 GPa hydrostatic pressure and the other under –1 GPa, both with heating from 0.01 K to 4000 K over 400 ps to explore the effect of pressure on the amorphous structure.

\subsubsection{Selection of crystalline and amorphous LCO structures for DFT calculations}
999 configurations from the MD trajectories are selected to proceed with the first-principles calculations. Fig. S13 shows the time evolution of the potential energy (eV/atom) for the simulations of cooling down an amorphous LiCoO$_{2}$ structure (density: 4.72 g/cm$^3$) to 0.01 K, and heating up the crystalline LiCoO$_{2}$ structure with the MP ID of 867664 to 3000 K. Take these two simulations as examples, for all the simulations starting with crystalline structures, 10 configurations with equal simulation steps between each other are selected from the whole trajectory, and for the regions where the phase transformation happens (where the slope of PE curve changes, indicated by the yellow region), 30 more configurations are selected. For the simulation of cooling down the amorphous LCO structure, 200 simulations are selected from the whole trajectory, and 30 more configurations are selected from the region where the glass transition happens. For the simulations of heating up the amorphous structure, 300 configurations are selected from the entire MD trajectory. The energies and densities of selected structures reproduced by DFT are shown in Fig. S1.

\subsection{Density functional theory calculations}
DFT+U calculations are performed using the generalized gradient approximation (GGA) with the Perdew–Burke–Ernzerhof (PBE) functional \cite{perdew1996generalized}, as implemented in the Vienna Ab initio Simulation Package (VASP) \cite{Kresse1993,Kresse1994,Kresse1996,Kresse1996a,hafner2008ab}. A plane-wave energy cutoff of 520 eV and an energy convergence threshold of 10$^{-4}$ eV/atom are used. A Hubbard U correction of 3.32 eV is applied to Co atoms \cite{dudarev1998electron}, and van der Waals (vdW) interactions are included via the DFT-D3 method \cite{Grimme2010}. These parameter choices are consistent with those used in the training of the PFP potential, as reported in \cite{takamoto2022towards}, and are adopted for all subsequent DFT calculations. Unless otherwise noted, calculations are performed without ionic relaxation to ensure consistency when comparing with PFP-based results. K-point meshes are generated using the pymatgen.io.vasp package (pymatgen.io.inputs.automatic\_density) \cite{Rong2016}, employing either Monkhorst–Pack or $\Gamma$-point sampling depending on the shape and size of the simulation cell. For calculations involving formation energies of LCO, both ionic and cell relaxations are allowed. 

Additionally, total energies are computed for various LCO structures reported in the MP database \cite{jain2020materials}, including MP\_853240, MP\_1097885, and MP\_1404711 (unit cells downloaded directly from MP) and MP\_753473 using a 2$\times$2$\times$1 supercell. For these structures, both ionic and cell relaxations are permitted. 

\subsection{Potential training using the NequIP/Allegro architecture}
We trained a neural network interatomic potential using Allegro \cite{musaelian2023learning}, an optimized variant of the NequIP framework \cite{batzner20223}. NequIP provides the general E(3)-equivariant graph-neural-network formalism, while Allegro implements a strictly local, highly scalable architecture that operates on ordered pairs of neighboring atoms. By constructing rich local descriptors from both invariant (scalar) and equivariant (tensor) latent features and propagating them through multiple interaction layers, the Allegro model achieves high accuracy for energies, forces, and stresses while remaining efficient on modern CPU+GPU hardware. At each layer, Allegro computes weighted tensor products between the current pairwise features and an embedded representation of the central atom’s local environment, where the weights themselves are learned from scalar features using another MLP. Scalar outputs from these tensor products are reintegrated into the invariant space and updated via residual connections, while equivariant outputs are linearly mixed within their symmetry channels. The final pairwise energy is predicted from the last layer’s scalar features using an output MLP. By avoiding atom-centered message passing and relying solely on local interactions, Allegro achieves linear scaling with system size and enables efficient parallelization, making it well-suited for large-scale molecular dynamics simulations.

Several key hyperparameters are systematically explored during model training:

\begin{itemize}
\item $r\_max$: the radial cutoff (in length units), which influences how far atomic interactions are considered. Higher values improve accuracy but increase training time. Values tested: 3.0, 4.0, 5.0, 6.0, 7.0.
\item $l\_max$: the maximum angular momentum quantum number for the spherical harmonics embedding. Larger values allow richer angular features at the cost of computational efficiency. Values tested: 0, 1, 2.
\item $parity$: determines the symmetry group—o3\_full, o3\_restricted, or so3—with descending expressive power.
\item $num\_layer$: the number of tensor product layers. Values tested: 1, 2, 3.
\item $env\_embed\_multiplicity$: the dimensionality of features in each tensor product layer. Values tested: 16, 32, 64, 128.
\item $two\_body\_latent\_mlp\_latent\_dimensions$: defines hidden-layer sizes for the two-body embedding MLP. The first layer is chosen from 16, 32, 64, or 128; each subsequent layer is set to twice the size of the previous one. 
\item $latent\_mlp$: controls hidden-layer sizes for the latent MLP. The latent MLP is one layer less than the two-body MLP with dimensions matching the final layer of the two-body MLP.
\item $PolynomialCutoff\_p$: the exponent in the polynomial envelope function, influencing how quickly the function decays with interatomic distance, Values tested: 6, 24, 48, based on prior work by Klicpera et al. \cite{gasteiger2020directional} and Musaelian et al. \cite{musaelian2023learning}.  
\noindent
\end{itemize}

Random combinations of these hyperparameters are tested, with other model parameters held fixed. The two-body latent MLP and later latent MLPs employ Sigmoid-weighted Linear Units (SiLU) activation functions \cite{Hendrycks2016}, while the environment embedding MLP is implemented as a single linear layer without nonlinearity. The final per-edge energy MLP includes one hidden layer of dimension 128 and no activation. All MLPs are initialized using a uniform distribution with unit variance. Training uses a joint loss function incorporating energy, force, and virial stress components, each with a weight of 1 due to the normalization of energy and stress terms. Optimization is performed using the Adam optimizer \cite{Kingma2014} in PyTorch \cite{Paszke2019} ($\beta1$ = 0.9, $\beta2$ = 0.999, $\epsilon$ = 10$^{-8}$) with no weight decay. A learning rate of 0.001 and batch size of 1 are used. Learning rate decay is controlled by a plateau scheduler with a patience of 25 epochs and a decay factor of 0.5. An exponential moving average with a decay factor of 0.99 is applied for model evaluation and selection. Training is terminated upon reaching any of the following: (a) a maximum training wall time of 12 hours, (b) 100,000 epochs, (c) no improvement in validation loss for 50 consecutive epochs, or (d) the learning rate dropping below 1×10$^{-5}$. All models are trained using float64 precision. 

The complete dataset consists of 999 structures, with 99 randomly selected for evaluation purposes only. The remaining 900 structures are split into 800 for training and 100 for validation. To improve generalization, the dataset was reshuffled at the end of each training epoch. 

\subsection{Calculations of LCO properties using the trained NNP}
\noindent\textbf{Lattice parameters, volume and density} Lattice constants $a$ and $c$ for structures at 0 K are obtained by structure optimization of the 360-atom LCO using the conjugate gradient method \cite{polyak1969conjugate} with box relaxation. The density is then calculated from the relaxed cell volume and the molar mass of LiCoO$_{2}$ (97.87 g/mol). The values for 300 K are obtained by maintaining the structure at 300 K for 15 ps with atoms being relaxed under the isothermal-isobaric ($NpT$) ensemble using the Nose-Hoover thermostat, with the temperature adjusted every 100 time steps with one integration step of 0.1 fs, and the pressure along each dimension adjusted every 1000 time steps with one integration step of 0.1 fs. The lattice constants and volume are then calculated by time-averaging over 1000 configurations for the last 5 ps. The same simulation conditions are applied to the calculations of the coefficient of thermal expansion.

\noindent\textbf{Young's modulus} To obtain the Young's modulus $E_{(100)}$ and $E_{(001)}$, strains ranging from -1\% to 1\% are applied to the box along the X-axis and Z-axis, respectively. The stress-strain curves are then fitted using linear regressions to obtain the Young's moduli (see Fig. S14(a)). To obtain the bulk modulus, strains ranging from -1\% to 1\% are applied to all axes of the simulation box, and fitting the volume-pressure curve using a linear regression (see Fig. S14(b)). 

\noindent\textbf{Coefficient of thermal expansion} 
Volumes at different temperatures can be obtained alongside with the phonon calculations. The temperature-volume curve is then fitted by an linear regression (see Fig. S14(c)), and the coefficient of thermal expansion at 300 K is calculated using the following equation,

\begin{equation}
    \alpha_{V}=\frac{1}{V}(\frac{\partial V}{\partial T})_{p}
    \label{cte}
\end{equation}

\noindent\textbf{Bulk modulus} The bulk modulus is then calculated using Eq. (\ref{bulk}).
\begin{equation}
    K=-V(\frac{d P}{dV})
    \label{bulk}
\end{equation}

\noindent\textbf{$\Gamma$ point vibrational modes} A smaller rhombohedral supercell (3$\times$3$\times$3) is employed for calculating the $\Gamma$ point vibrational modes to reduce the computational cost. These calculations are carried out using the Phonon3py module \cite{togo2023implementation} integrated into the ASE framework \cite{larsen2017atomic}. This module, which builds on the same approach as the Phonon module, estimates phonon relaxation times and computes thermal conductivity by solving both the phonon Boltzmann equation and the Wigner transport equation. 

\noindent\textbf{Isobaric heat capacity} The lattice heat capacity was computed using the Phonon module interfaced with ASE and Phonopy \cite{larsen2017atomic}. Harmonic phonon frequencies were obtained from force-constant matrices generated by the small-displacement method \cite{alfe2009phon} on a dense \textbf{q}-point mesh. For each volume in a set of strained cells, the phonon density of states was used to evaluate the Helmholtz free energy, entropy, and constant-volume heat capacity $C_v$(T). These quantities were then passed to the PhonopyQHA module, which performs a quasi-harmonic analysis to determine the equilibrium volume as a function of temperature and to compute the constant-pressure lattice heat capacity $C_p$(T). The rhombohedral primitive cell of the layered LiCoO$_{2}$ structure is used (see Fig. S15), as ASE correctly identifies the corresponding space group when this cell is used. Structural relaxation of the primitive cell is performed using the fast inertial relaxation engine (FIRE) algorithm \cite{bitzek2006structural}, with convergence criteria requiring maximum atomic forces below 1$\times$10$^{-4}$ eV/\text{\AA}, and minimal residual hydrostatic pressure. A 6$\times$6$\times$6 supercell is constructed based on the relaxed cell for subsequent phonon calculations. The high-symmetry Brillouin zone path ($\Gamma$-X-Y-$\Gamma$-Z-R-$\Gamma$-T-U-$\Gamma$-V) is generated using the Seek-path Module \cite{hinuma2017band,togo2024spglib}, and phonon properties are sampled over a 20$\times$20$\times$20 q-point mesh. The temperature range for the calculations spans from 0 K to 1400 K in 10 K increments. This calculation also yields the temperature dependence of both volume and bulk modulus, which are summarized in Table S1 to compare with the values obtained by LAMMPS.

\noindent\textbf{Thermal conductivity} Thermal conductivity was evaluated using two approaches. First, the anisotropic lattice thermal conductivity was computed with phono3py by solving the linearized phonon Boltzmann transport equation within the single-mode relaxation time approximation (RTA). A hexagonal 3$\times$3$\times$1 supercell was constructed so that the simulation cell Z-axis coincides with the crystallographic $c$-axis, enabling separate evaluation of in-plane ($\kappa_{ab}$) and out-of-plane ($\kappa_{c}$) components (see Fig. S15). Harmonic and third-order force constants were obtained using finite displacements, and a 19$\times$19$\times$19 q-point mesh was employed for Brillouin-zone sampling. Calculation results are plotted against DFT data in Fig. S7(a).

The second approach is the Green--Kubo formalism, implemented in LAMMPS 
following Lee et al.~\cite{Lee2024}. Within this framework, the thermal 
conductivity tensor is defined as:
\begin{equation}
\label{eq:gk_kappa}
    \kappa_{\alpha\beta}(T) = \dfrac{1}{k_\mathrm{B} T^2 V} 
    \lim_{\tau\to\infty} \int_{0}^{\tau} dt \, 
    \langle j_\alpha(t) \cdot j_\beta(0) \rangle_{T},
\end{equation}
where $k_\mathrm{B}$ is the Boltzmann constant, $T$ the temperature, $V$ 
the volume, $j_\alpha(t)$ the $\alpha$-th Cartesian component of the 
macroscopic heat flux, and $\langle j_\alpha(t) \cdot j_\beta(0) \rangle_{T}$ 
the heat flux autocorrelation function (HFACF), with $\langle \cdot \rangle_T$ 
denoting the ensemble average over time and over independent MD trajectories.

For each temperature, lattice parameters were first obtained from $NpT$ 
simulations using a timestep of 0.1 fs. The system was then thermalized in the $NVT$ ensemble for 100~ps, 
after which production runs were performed in the $NVE$ ensemble with the heat 
flux recorded throughout. The ensemble average in Eq.~\eqref{eq:gk_kappa} is 
obtained from 20 independent MD trajectories with different initial velocities. 
The trajectory length was adjusted for each system to ensure convergence of the 
HFACF, ranging from 1.6 to 2.4~ns.

Green--Kubo calculations require sufficiently large simulation cells and long 
trajectories to obtain converged thermal conductivities. To assess finite-size 
effects, we computed the lattice thermal conductivity of rhombohedral LCO at 
300~K using supercells containing 360, 2880, and 9720 atoms. The running thermal conductivities averaged over these trajectories are shown in Fig.~S16, with shaded bands indicating the standard error. While the 360-atom cell underestimates both in-plane and cross-plane conductivities, the 2880- and 9720-atom results converge to similar plateau values for correlation times $\tau \gtrsim 25$~ps, with overlapping error bands. The largest (9720-atom) system, however, exhibits larger statistical noise and is significantly more computationally demanding. We therefore adopt the 2880-atom supercell as a size-converged and computationally efficient choice, and use it for Green--Kubo calculations at 100, 200, 900, and 1500~K (Fig.~S17).

\subsection{Optical Property Measurements}

The wavelength-dependent optical properties of as-deposited and annealed LCO thin films were characterized by UV-Vis-NIR spectrophotometry. Films were deposited by RF magnetron sputtering on borosilicate glass substrates (25 $\times$ 25 mm$^2$) following the procedure described in \cite{chen2021flash}, and the film thickness was determined to be 700 nm by stylus profilometry.
Transmittance (T) and reflectance (R) spectra were measured at room temperature in ambient atmosphere using a Shimadzu UV-3600 spectrophotometer over the wavelength range 300-1500 nm with a spectral resolution of 2 nm. The wavelength-dependent absorptance A($\lambda$) was then obtained from the measured spectra via
\begin{equation}
A(\lambda) = 1 - T(\lambda) - R(\lambda)
\end{equation}

The absorption coefficient ($\alpha$, also referred to as the attenuation coefficient) was extracted from the optical measurements using:
\begin{equation}
\alpha(\lambda) = \frac{1}{d} \ln \left[ \frac{(1-R(\lambda))^2}{T(\lambda)} \right]
\end{equation}
where $d$ is the film thickness. 


\section{Acknowledgements}

This research was supported by the Swiss National Science Foundation (SNSF) under Project No. 200021\_219741 and by NCCR MARVEL, a National Centre of Competence in Research, funded by the Swiss National Science Foundation (grant number 205602). The authors thank the Swiss National Supercomputing Centre (project no. lp126) and GENCI-IDRIS (grant no. 2025-A0190914626) for computing resources.

\section{Data availability}
Data supporting the findings of this study are openly available at the following URL/DOI: \url{https://doi.org/10.24435/materialscloud:3e-w9}
\section{Author Contributions}
Y. H. conducted main calculations, generated and analyzed the data, interpreted the results, and prepared the first draft. B.S. conducted selected calculations, generated and analyzed associated data, and contributed to interpretation of the results. W.V. performed experimental measurements and analyzed the experimental data. Y.R. secured funding and supported project administration. V.T. conceptualized the research, and oversaw the collaborative project. All authors contributed to writing and editing the manuscript.

\section{Competing Interests}
The authors declare no competing interests.

\section{Additional information}
Supplementary Information accompanies this article. It includes:
(i) details of thermal cycling simulations with the universal Matlantis potential and its performance against DFT on structures sampled from MD trajectories;
(ii) the development and hyperparameter optimization of the Allegro potential, its accuracy relative to DFT, and scalability benchmarks on modern supercomputing infrastructure;
(iii) additional validation of the trained Allegro potential against known structural, elastic, vibrational, and thermodynamic properties of LCO;
(iv) the methodology used to estimate grain-size–dependent thermal conductivity;
(v) a more detailed discussion of the physical origin of enhanced optical absorption in amorphous LCO;
(vi) full specifications of the multiphysics photothermal simulations, including material parameters, boundary conditions, and lamp pulse inputs; and
(vii) supplementary figures supporting the main text.

\bibliographystyle{ieeetr} 
\bibliography{sn-bibliography}

\end{document}